\documentclass{article}
\usepackage{graphicx}
\usepackage{epsfig}
\usepackage{amsfonts}
 \usepackage{amssymb}
\usepackage{enumitem}

\date{\today}

\newcommand{\insertplot}[5]{\begin{figure}
 \hfill\hbox to 0.05in{\vbox to #5in{\vfill
 \inputplot{#1}{#4}{#5}}\hfill}
 \hfill\vspace{-.1in}
 \caption{#2}\label{#3}
 \end{figure}}
 \newcommand{\inputplot}[3]{
 \special{ps: plotfile #1}
}
\newcounter{fig}

\newcommand{\ee}{\end{equation}}
\newcommand{\eea}{\end{eqnarray}}
\newcommand{\be}{\begin{equation}}
\newcommand{\bea}{\begin{eqnarray}}

\usepackage[usenames,dvipsnames]{color}

\begin{document}

\title{ \bf 
Higher dimensional black hole scalarization}
 \vspace{1.5truecm}
\author{
  Dumitru Astefanesei$^1$, Carlos Herdeiro$^2$, Jo\~ao Oliveira$^2$
and  Eugen Radu$^{2}$  
\\
\\
%
{\small $^1$ Pontificia Universidad Cat\'olica de
	Valpara\'\i so, Instituto de F\'{\i}sica},
 \\ 
{\small Av. Brasil 2950, Valpara\'{\i}so, Chile} 
\\
{\small 
$^2$
Departamento de Matem\'atica  da Universidade de Aveiro and}
\\
{\small 
 Center for Research and Development in Mathematics and Applications (CIDMA),
}
\\
{\small 
 Campus de Santiago, 3810-183 Aveiro, Portugal
}
}

\date{July 2020}

\maketitle 

\begin{abstract}
In the simplest scalar-tensor theories, wherein the scalar field is non-minimally coupled to the Ricci scalar, spontaneous scalarization of electrovacuum black holes (BHs) does not occur. This ceases to be true in higher dimensional spacetimes, $d>4$. We consider the scalarization of the higher dimensional Reissner-Nordstr\"om BHs in scalar-tensor models and provide results on the zero modes for different $d$, together with an explicit construction of the scalarized BHs in $d=5$, discussing some of their properties. We also observe that a conformal transformation into the Einstein frame maps this model into an Einstein-Maxwel-scalar model, wherein the non-minimal coupling occurs between the scalar field and the Maxwell invariant (rather than the Ricci scalar), thus relating the occurence of scalarization in the two models. Next, we consider the spontaneous scalarization of the Schwarzschild-Tangherlini BH in extended-scalar-tensor-Lovelock gravity in even dimensions. In these models, the scalar field is non-minimally coupled to the $(d/2)^{th}$ Euler density, in $d$ spacetime dimensions. We construct explicitly examples in $d=6,8$, showing the properties of the four dimensional case are qualitatively generic, but with quantitative differences. We compare these higher $d$ scalarized BHs with the hairy BHs in shift-symmetric Horndeski theory, for the same $d$, which we also construct.
\end{abstract}
  
	
\section{Introduction}
Spontaneous scalarization triggered by strong gravity effects emerges in some classes of scalar-tensor models. 
This phenomenon could provide a smoking gun for scalar-tensor theories and may be interpreted as a strong gravity phase transition.  Here, for concreteness, we shall be considering models described by the generic $d$-dimensional action
\be
\label{actiongen}
\mathcal{S}=- \frac{1}{16 \pi}\int d^d x \sqrt{-g} 
\left\{
[1-\alpha_{\rm st}f_{\rm st}(\phi)]R  
-\frac{1}{2}\partial_\mu  \phi \partial^\mu \phi 
+\alpha_{\rm L} f_{\rm L}(\phi) \mathcal{L}_{(p)}-\alpha_{\rm emg}f_{\rm emg}(\phi)F_{\mu\nu}F^{\mu\nu} -\mathcal{L}_{\rm mat}
\right\} \ .
\ee
$R$ is the Ricci scalar of the $d$-dimensional spacetime metric, $g_{\mu\nu}$, with determinant $g$; $\phi$ is a real scalar field, with a canonical kinetic term; $F_{\mu\nu}$ is the Maxwell 2-form; $\mathcal{L}_{\rm mat}$ is an unspecified matter Lagrangian and $\mathcal{L}_{(p)}$ is the $p^{\rm th}$ Euler density:
\begin{eqnarray}
\label{Lp-my}
\mathcal{L}_{(p)}\equiv \frac{(2p)!}{2^p}
\delta^{\mu_1}_{[\rho_1}\cdots \delta^{\mu_{2p}}_{\rho_{2p}]}
R_{\mu_1\mu_2}^{\phantom{\mu_1}\phantom{\nu_1}\rho_1\rho_2}\cdots R_{\mu_{2p-1}\mu_{2p}}^{\phantom{\mu_{2p-1}}\phantom{\mu_{2p}}\rho_{2p-1}\sigma_{2p}} \ ,
\end{eqnarray}
where $R_{\mu\nu}^{\phantom{\mu}\phantom{\nu}\alpha\beta}$ is the Riemann tensor. The functions $f_{\rm st}(\phi),f_{\rm L}(\phi)$ and $f_{\rm emg}(\phi)$ are three unspecified non-minimal coupling functions that, when appropriately chosen, lead to spontaneous scalarization, in each case triggered by a different source; the strength of each of the effects is controlled by the coupling \textit{constants}, $\alpha_{\rm st},\alpha_{\rm L},\alpha_{\rm emg}$. In geometrized units, $\alpha_{\rm st}$ and $\alpha_{\rm emg}$ are dimensionless,  whereas $\alpha_{\rm L}$ has dimensions of length squared. 

The original scalarization mechanism~\cite{Damour:1993hw} was proposed in $d=4$ scalar-tensor theories with 
\be
\alpha_{\rm st}\neq 0 \ , \qquad \alpha_{\rm L}=0=\alpha_{\rm emg} \ .
\label{guise1}
\ee
Scalar-free objects that may become scalarized must have $R\neq0$. This is the case, $e.g.$ of neutron stars, but not of the electrovacuum black holes (BHs), which are immune to scalarization in this framework. 

More recently, it was observed~\cite{Silva:2017uqg,Doneva:2017bvd,Antoniou:2017acq} that the scalarization of vacuum BHs could occur in $d=4$ \textit{extended} scalar-tensor theories with 
\be
\alpha_{\rm L}\neq 0 \ , \qquad \alpha_{\rm st}=0=\alpha_{\rm emg} \ .
\label{guise2}
\ee
Scalarization now requires a non-vanishing  Gauss-Bonnet (GB) invariant, which holds for vacuum  BHs, whose Kretschmann  scalar is non-vanishing, despite being Ricci flat\footnote{Scalarization should occur also in Einstein-Chern-Simons models
with a suitable coupling between the scalar field and the Pontryagin density.
No static scalarized solutions, however, have yet been studied, the only case investigated so far being the NUT generalization of the Schwarzschild BH \cite{Brihaye:2018bgc}. }. Scalarized solutions have been constructed in these models, but a dynamical study of the full scalarization process, from the initial trigger around a vacuum BH until the settling into a scalarized BH is still lacking. 

In parallel, it was observed~\cite{Herdeiro:2018wub} that scalarization of $d=4$ electrovacuum BHs occurs in Einstein-Maxwell-scalar models which have 
\be
\alpha_{\rm emg}\neq 0 \ , \qquad \alpha_{\rm st}=0=\alpha_{\rm L} \ .
\label{guise3}
\ee
In this guise of spontaneous scalarization, the trigger is a non-vanishing Maxwell invariant $F_{\mu\nu}F^{\mu\nu}$ and thus the phenomenon  needs not gravity; moreover, in this case it was possible to establish dynamically that the  scalarization of electrically charged electrovacuum BHs indeed occurs, when these BHs have sufficiently high charge, and that the evolution settles into the scalarized solutions that can be constructed as stationary states of the field equations~\cite{Herdeiro:2018wub,Fernandes:2019rez}.

\medskip

These three guises of spontaneous scalarization have been considered in $d=4$. Considering $d\neq 4$ raises interesting questions, which also address the universality of the phenomenon. Firstly, for the $d>4$ electrovacuum BHs, $i.e.$ the higher dimensional generalizations of the Reissner-Nordstr\"om (RN) solution (see $e.g.$~\cite{Chamblin:1999hg}), $R=0$ ceases to hold, since classical electromagnetism is only conformally invariant in $d=4$. Thus, higher dimensional charged BHs can be scalarized in the original scalar-tensor models~(\ref{guise1}), with $\mathcal{L}_{\rm mat}=F_{\mu\nu}F^{\mu\nu}$. Here, we shall show this indeed occurs and construct explicit scalarized RN BHs in these models.\footnote{Taking into account quantum corrections, electrovacuum BHs can also become scalarized in the original scalar-tensor model~\cite{Herdeiro:2019yjy}.}

Secondly, one may inquire if there is anything special about the scalarization in $d=4$ extended scalar-tensor models~(\ref{guise2}), or if similar scalarized BHs occur in $d\neq 4$. We shall show that, indeed, the phenomenon is universal, and the properties of the higher dimensional scalarized BHs, using the appropriate Euler density, are similar to the ones of the four dimensional model with the GB term. 

Finally, the simultaneous  consideration of these three different guises of spontaneous scalarization raises the following question: models~(\ref{guise1})  and~(\ref{guise3}) can be mapped into one another (for particular couplings) via a conformal transformation; how does this mapping allow relating scalarized solutions of both models in $d>4$? Here, we shall provide the explicit mapping and exemplify how information can be extracted from it.

\medskip
This paper is organized as follows. In Section~\ref{sec2} we review the basic ingredients for spontaneous scalarization. In Section~\ref{sec3} we address the scalarized electrovacuum BHs in scalar-tensor $d>4$ models, constructing the zero modes for general $d$ and the scalarized BHs for the simplest coupling function allowing scalarization in $d=5$, exhibiting some of their properties. We also address the mapping into the Einstein frame and the relation with Einstein-Maxwell-scalar models. In Section~\ref{sec4} we consider higher $d$ extended scalar-tensor theories, where the scalar field non-minimally couples to the appropriate Lovelock density. Again we construct the scalarized BHs for $d=6,8$ (besides $d=4$) and discuss some of their properties. We also compare them with the hairy BHs in shift-symmetric Horndeski for the same $d$ emphasizing some of the differences between the two models. We conclude with a summary and discussion in Section \ref{sec5}.

\section{A scalarization primer}
\label{sec2}
Scalarized BHs occur in the family of models~(\ref{actiongen}), co-existing with scalar-free electrovacuum BHs for appropriate choices of the coupling functions $f_{\rm st}(\phi),f_{\rm L}(\phi)$ and $f_{\rm emg}(\phi)$. These solutions violate  the scalar no-hair theorems (see $e.g.$~\cite{Herdeiro:2015waa})
due to the 
presence of a tachyonic  {\it effective mass} term
in the
scalar field action.
To understand the source of the effect, consider the scalar sector of the action~(\ref{actiongen}), which reads
\begin{eqnarray}
\label{actionS}
\mathcal{S}_\phi= -\int  d^d x \sqrt{-g} 
\left[
 \frac{1}{2}\partial_\mu \phi \partial^\mu \phi
+ \alpha f(\phi) {\cal I}
\right] \ ,
\end{eqnarray}
where $f(\phi)$ and $\alpha$ stand for any one of the three coupling constants and coupling functions in~(\ref{actiongen}) 
and 
${\cal I}$ is the source term, which in~(\ref{actiongen})  would be one the three choices
\be
\mathcal{I}=\left\{ R, -\mathcal{L}_{(p)}, F_{\mu\nu}F^{\mu\nu}\right\} \ .
\ee 
Then,  
the scalar field equation of motion 
reads
\begin{eqnarray}
\label{eq-phi}
\Box \phi=\alpha \frac{d f}{d \phi}{\cal I} \ .
\end{eqnarray}

The phenomenon of ``spontaneous scalarization" requires the following ingredients:
\begin{itemize}
\item[i)]
There exists a \textit{scalar-free} solution with $\phi=\phi_0$.
For $\mathcal{I}\neq 0$ eq. (\ref{eq-phi}) implies the coupling function should satisfy the condition
\begin{eqnarray}
\label{condx}
\frac{d f}{d \phi}\Big |_{\phi=\phi_0}=0 \ .
\end{eqnarray}
One may set $\phi_0=0$ via a field redefinition.
Thus electrovacuum BHs plus a vanishing scalar field solve the model.
\item[ii)]
The scalar-free solution suffers a tachyonic instability triggered by linear scalar perturbations.
 For a small scalar field $\phi=\delta\phi$, linearizing (\ref{eq-phi}) yields
\begin{eqnarray}
\label{eq-phi-small}
(\Box-\mu_{\rm eff}^2)\delta\phi =0 \ , \qquad {\rm where} \qquad \mu_{\rm eff}^2\equiv  \frac{d^2 f}{d \phi^2}\Big |_{\phi=0} \mathcal{I} \ .
\end{eqnarray} 
Without loss of generality we choose
${d^2 f}/{d \phi^2}\Big |_{\phi=0} $
to be strictly positive.
Then the tachyonic condition
$\mu_{\rm eff}^2<0$
implies  
\begin{eqnarray}
\label{conds}
\mathcal{I}<0~,
\end{eqnarray}
must hold \textit{for some region} outside the horizon.
\item[iii)]
A second set of solutions exist, with a nontrivial scalar field, the \textit{scalarized BHs}. 
These solutions are continuoulsy connected with the scalar-free set, approaching it as $\phi\rightarrow 0$. In this limit the scalar field becomes a \textit{scalar cloud} or \textit{zero mode}. Although the quantitative properties of these solutions depend on the choice of the coupling function, qualitative properties are not so sensitive, as long as the condition (\ref{condx}) is satisfied. 
\end{itemize}

\section{Scalarized electrovacuum BHs in $d>4$ scalar-tensor  models}
\label{sec3}

\subsection{The framework }
 For our first analysis we consider a scalar-tensor model, with the matter Lagrangian describing classical electromagnetism. Thus, we take 
(\ref{actiongen}) with (\ref{guise1}) and $\mathcal{L}_{\rm mat}=F_{\mu\nu}F^{\mu\nu}$. Moreover, we take the simplest coupling function allowing for spontaneous scalarization:
\be
f_{\rm st}(\phi)=\phi^2 \ ,
\ee
and for ease of notation we drop the subscript label in the coupling constant: $\alpha_{\rm st}\rightarrow \alpha$.
As such, the action of the model reads
\be
\label{actionEM}
\mathcal{S}=- \frac{1}{16 \pi}\int d^d x \sqrt{-g} 
\left\{
(1-\alpha \phi^2 )R  
-\frac{1}{2}\partial_\mu  \phi \partial^\mu \phi 
-F_{\mu\nu}F^{\mu\nu} 
\right\} ;
\ee
observe that the scalar and electromagetic fields interact only indirectly, via the backreaction on the spacetime metric.

Restricting to spherically symmetric configurations,
we consider a metric ansatz 
in Schwarzschild-like coordinates, together with a scalar  field and electric potential 
which depend on the radial coordinate only,
\begin{eqnarray}
\label{metric-scalar}
ds^2=-N(r)\sigma^2(r)dt^2+\frac{dr^2}{N(r)}+r^2 d\Omega_{d-2}^2~, \qquad \phi \equiv \phi(r)\ , \qquad A=V(r) dt \ .
\end{eqnarray}
The coordinates $(r,t)$ possess the usual meaning and $d\Omega_{d-2}^2$  is the line element on the unit $(d-2)$-sphere. This ansatz results in the following  equations (where the ``prime" denotes radial derivatives):
\begin{eqnarray} 
&&
(d-2) N'
-(d-2)(d-3) \frac{(1-N)}{r}
+\frac{1}{2}r N\phi'^2
+\frac{2r V'^2}{\sigma^2}
\nonumber
\\
&&
+4\alpha 
\left\{
rN\phi \phi''
+r N\phi'^2 
+\frac{1}{2}\phi\phi' \left[r N'+2(d-2) N\right]
+\frac{d-2}{4} \phi^2 \left[N'-\frac{d-3}{r}(1-N)\right]
\right\}=0 \ ,
\label{st1}
\\
&&
\sigma'-\frac{r \sigma \phi'^2}{2(d-2)}
-\frac{\alpha}{(d-2)}
\left\{
\phi \sigma' [(d-2)\phi+2r\phi']
-2r \sigma
(
\phi'^2+\phi \phi''
)
\right\}=0 \ ,
\\
&&
(r^{d-2}N\sigma \phi')'
+2\alpha \phi
\left\{
r^{d-3}[\sigma (rN''+(d-2)N')
+\sigma' (3r N'+2(d-2) N)
+2r N\sigma'']-{\cal L}_{E}
\right\}
=0 \ , \qquad 
\\
&&
V''+
\left[
\frac{(d-2)}{r}-\frac{\sigma'}{\sigma}
\right]V'=0 \ .
\label{st2}
\end{eqnarray}
These equations can also be derived from the effective Lagrangian
\begin{eqnarray}
{\cal L}_{\rm eff}={\cal L}_{\rm E}+{\cal L}_{s}+{\cal L}_{\rm M}+\mathcal{L}_{\rm R} \ ,
\end{eqnarray}
where  
\begin{eqnarray} 
\label{lels}
 {\cal L}_{\rm E}=(d-2)r^{d-4}\sigma 
\left\{
r N'+(d-3)(N +1)+2N r \frac{\sigma'}{\sigma}
\right\} \ , \qquad 
 {\cal L}_{s}=-\frac{1}{2}N\sigma r^{d-2}\phi'^2\ , 
\end{eqnarray}
and
\begin{eqnarray} 
\nonumber
 {\cal L}_{\rm M}=\frac{2r^{d-2}V'^2}{\sigma}\ ,
\qquad 
 {\cal L}_{\rm R}= -\alpha \phi 
\left[
\phi {\cal L}_{E}+2r^{d-2}\phi'(\sigma N'+2N \sigma')
\right] \ .
\end{eqnarray}
The equation  for the electric potential  possesses the first integral, which, for convenience we write as
\begin{eqnarray}
\label{first-int}
V'=  \frac{(d-3)Q_0 \sigma}{r^{d-2}} \  ,
\end{eqnarray}
where $Q_0$ is an integration constant fixing the electric charge.

We are interested in BH solutions,
with an horizon at $r=r_h>0$.
Restricting to non-extremal configurations,
the solutions possess a near horizon expansion with the first terms being
\begin{eqnarray}
\label{horizon1}
&&
N(r)=N_1(r-r_h)+\dots\ , \qquad 
\sigma(r)=\sigma_0 + \sigma_1 (r-r_h)+\dots\ , 
\\
&&
\nonumber
\phi(r)=\phi_0 + \phi_1 (r-r_h)+\dots \ , \qquad 
V(r)=v_1 (r-r_h)+\dots \ , 
\end{eqnarray} 
which contains two essential
parameters
$\phi_0$
and 
$\sigma_0$ (the remaining ones are determined in terms of these).

The approximate form of the solutions in the far field reads
\begin{equation}
\label{inf1}
N(r)=1-\frac{m}{r^{d-3}}+\dots\ , ~ \phi(r)=\frac{Q_s}{r^{d-3}}+\dots\ , ~
V(r)=\Phi-\frac{Q_0}{r^{d-3}}+\dots \ , ~
\sigma(r)=1-\frac{[d-3-4\alpha (2d-5)]Q_s^2}{4(d-2)r^{2(d-3)}}\ +\dots \ .
\end{equation}
Apart from $m$ and $Q_0$,
the essential parameters here are
$\Phi$ (the electrostatic potential)
and
$Q_s$
(the scalar `charge'). Thus, the data at infinity is specified by the ADM mass $M$,
electric charge $Q$,  electrostatic potential $\Phi$
and scalar 'charge' $Q_s$,
which are read off from the far field asymptotics (\ref{inf1}), where the physical $M,Q$ relate to the parameters $m,Q_0$ as
 \begin{eqnarray}
\label{QMgen}
M=\frac{(d-2)V_{(d-2)}}{16 \pi}m \ , \qquad Q=\sqrt{2(d-2)(d-3)}\frac{ V_{(d-2)}}{8 \pi}Q_0 \ ,
\end{eqnarray}
and $V_{(d-2) }$ is the area of a  $(d-2)$-sphere. 

The  horizon data, on the other hand,  is the Hawking temperature $T_H$ and the event horizon area $A_H$, which are given by
 \begin{eqnarray} 
\label{THAH}
 T_H=\frac{(d-3)}{4\pi r_h}
\left(1-\frac{[(d-3)Q_0]^2}{r_h^{2(d-3)}}
\right)\ , \qquad 
A_H=V_{(d-2)}r_h^{d-2}\ , 
\end{eqnarray}
together with the value at 
$r=r_h$
of the scalar field
$\phi=\phi_0$.

We also define the reduced electric charge, horizon area and temperature as
\begin{eqnarray}
\label{scale1}
q\equiv \frac{Q}{M}\sqrt{\frac{d-2}{2(d-3)}} \ , \qquad a_H\equiv \frac{A_H}{ M^{\frac{d-2}{d-3}}}c_a \ , \qquad t_H\equiv  T_H  M^{\frac{1}{d-3}}c_t \ ,
\end{eqnarray}
where  
\begin{eqnarray}
\label{scale2}
c_a= \frac{V_{(d-2)}^{\frac{1}{d-3}}(d-2)^{\frac{d-2}{d-3}}}
{(16\pi)^{\frac{d-2}{d-3}}} \ , \qquad 
c_t=\frac{2^{\frac{2(d-1)}{d-3}} \pi^{\frac{d-2}{d-3}} }{(d-3)(d-2)^{\frac{1}{d-3}}V_{(d-2)}^{\frac{1}{d-3}}}  \ .
\end{eqnarray}

The scalar-free solution of the model\footnote{Here we follow the conventions
used for this solution in \cite{Chamblin:1999hg}.} 
is the RN
BH, which is specified by~(\ref{metric-scalar}) with $\phi=0$
and  
\begin{eqnarray}
\label{RN}
 N(r)=1-\frac{m}{r^{d-3}}+\frac{2(d-3)}{(d-2)}\frac{Q_0^2}{r^{2(d-3)}} \ , \qquad \sigma(r)=1\ , \qquad  
V(r)=\Phi- \frac{Q_0}{r^{d-3}} \ .
\end{eqnarray}
The RN BH possess an (outer) horizon at
$r=r_h$,
where $r_h$ is the largest (positive) solution of the equation $N(r_h)=0$. 
Working in a gauge with $V(r_h)=0$,
the constant $\Phi$ corresponds to the electrostatic potential,
$\Phi= \frac{Q_0}{r_h^{d-3}}.$
We also remark that for 
$d>4$,
the RN BH possesses a nonvanishing
Ricci scalar:
\begin{eqnarray} 
R=\frac{d-4}{d-2}F^2=-\frac{2(d-3)^2(d-4)Q_0^2}{(d-2)r^{2(d-2)}} \ .
\end{eqnarray}

\subsection{The zero mode for general $d$}
Let us start by treating the scalar field as a small perturbation around the $d$-dimensional RN background. This will allow us to compute the \textit{zero modes}: linear scalar field bound states that are supported by a discrete set of RN backgrounds. Zero modes define the onset of the scalarization instability and the bifurcation towards the new family of scalarized BHs.

Restricting to a spherically symmetric scalar field, 
the  equation for $\phi$
reads  
\begin{eqnarray}
\label{p2}
 (r^{d-2}N \phi')'+\frac{4\alpha (d-3)^2(d-4)}{d-2}\frac{Q_0^2}{r^{d-2}} \phi=0 \ .
\end{eqnarray}
One can see that $\phi$-coefficient in the  above equation
acts as an $r$-dependent \textit{effective} mass  
  for the perturbations, with the condition $\alpha>0$ being necessary for a tachyonic mass.

We are interested in solutions of the above equation which are regular for $r\geqslant r_h$
and vanish 
at infinity.
Remarkably,
one finds the following exact solution
\begin{eqnarray}
\label{ex1}
\phi(r)=P_u 
\left[
1-\frac{2}{1-\frac{r_h^{2(d-3)}}{\frac{2(d-3)Q_0^2}{d-2} }}
\left\{1-\left(\frac{r_h}{r}\right)^{d-3}\right\}
\right] \ , \qquad 
{\rm with} \qquad 
u\equiv \frac{1}{2}\left(-1+\sqrt{1-\frac{8\alpha (d-4)}{(d-3)}} \right) \ ,
\end{eqnarray}
$P_u$ being the Legendre function. 
One can show that, in general, the function
$\phi(r)$ 
approaches a constant \textit{non-zero} value as $r\to \infty$,
\begin{eqnarray}
\label{ex2}
\phi(r) \to 
{}_2F_1 
\left[
\frac{1}{2}\left(1-\sqrt{1-\frac{8\alpha (d-4)}{(d-3)}}\right) \ ,
\frac{1}{2}\left(1+\sqrt{1-\frac{8\alpha (d-4)}{(d-3)}}\right),
1; 
\frac{1}{ 1-\frac{r_h^{2(d-3)}}{\frac{2(d-3)Q_0^2}{(d-2)}}}
\right]+\mathcal{O}\left(\frac{1}{r}\right) \ .
\end{eqnarray} 
Thus finding the spherically symmetric zero mode
of the RN BH 
within the model (\ref{actionEM}), corresponding to a scalar field that vanishes asymptotically, 
reduces to a study of the zeros of the hypergeometric function
${}_2F_1 $.

\begin{figure}[h!]
\begin{center}
\includegraphics[width=0.55\textwidth]{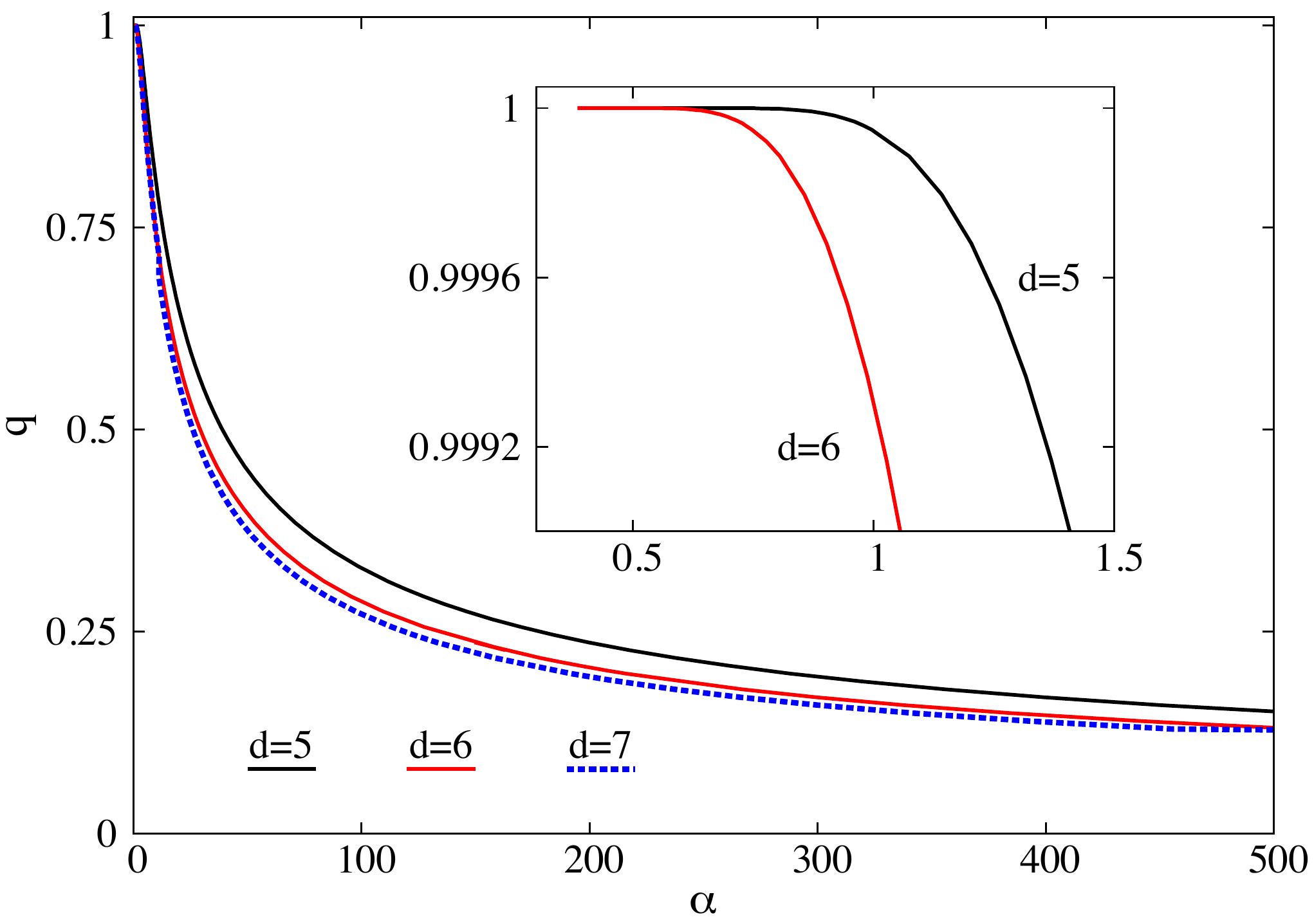}  
\caption{  
Existence lines for the $d=5,6,7$ RN BH in a $\alpha$ $vs.$ $q$  diagram. The inset zooms around the minimal value of $\alpha$.
}
\label{existence-lines}
\end{center}
\end{figure}

The \textit{existence line}, $i.e.$ the RN backgrounds that support scalar clouds, correspond to the set of values of $q\sim Q/M$,
 as a function of $\alpha$, 
for which $\phi(r)\to 0$ asymptotically. Such existence lines  
are illustrated in Fig.~\ref{existence-lines} for $d=5,6,7$. 
For any $d>4$, the solutions exist as long as the coupling constant is sufficiently large, $i.e.$ for 
\be
\frac{d-3}{8(d-4)}<\alpha<+ \infty \ ,
\label{con1s}
\ee
the minimal value corresponding to the $T_H\to 0$ limit of the RN background.

\subsection{An explicit construction: scalarized $d=5$ RN BHs}
 
The scalarized BH solutions obeying the asymptotic behaviours  (\ref{horizon1}) and (\ref{inf1})
are found numerically,  by using a standard ordinary differential equations (ODE) solver.
Here we shall report the $d=5$ case that we have studied more systematically.
We have also verified, however, the existence of scalarized solutions for $d=6$
and we conjecture the existence of such configurations
for any $d\geqslant 5$.
 \begin{figure}[h!]
\begin{center}
\includegraphics[width=0.45\textwidth]{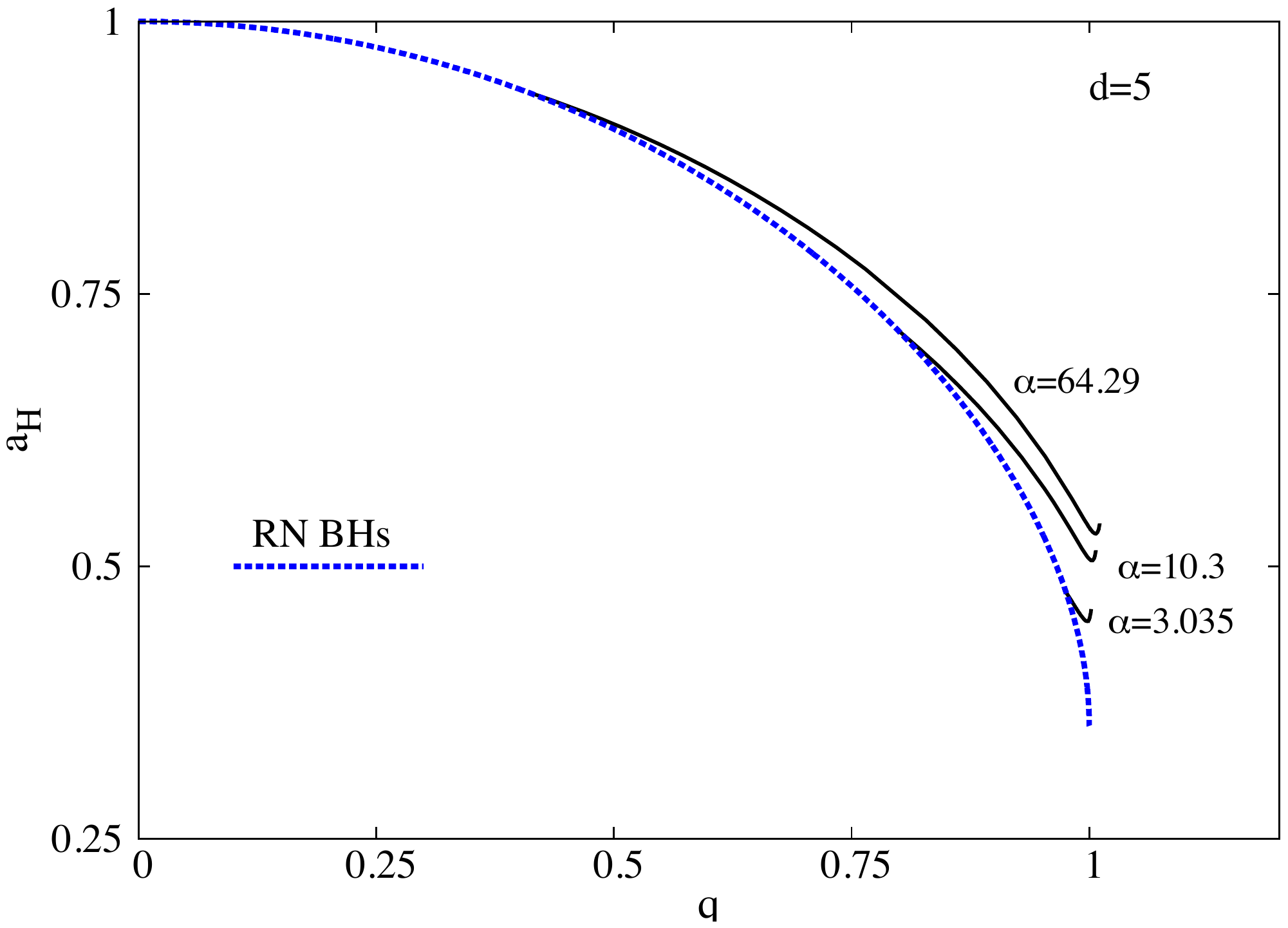}  
\ \ \
\includegraphics[width=0.45\textwidth]{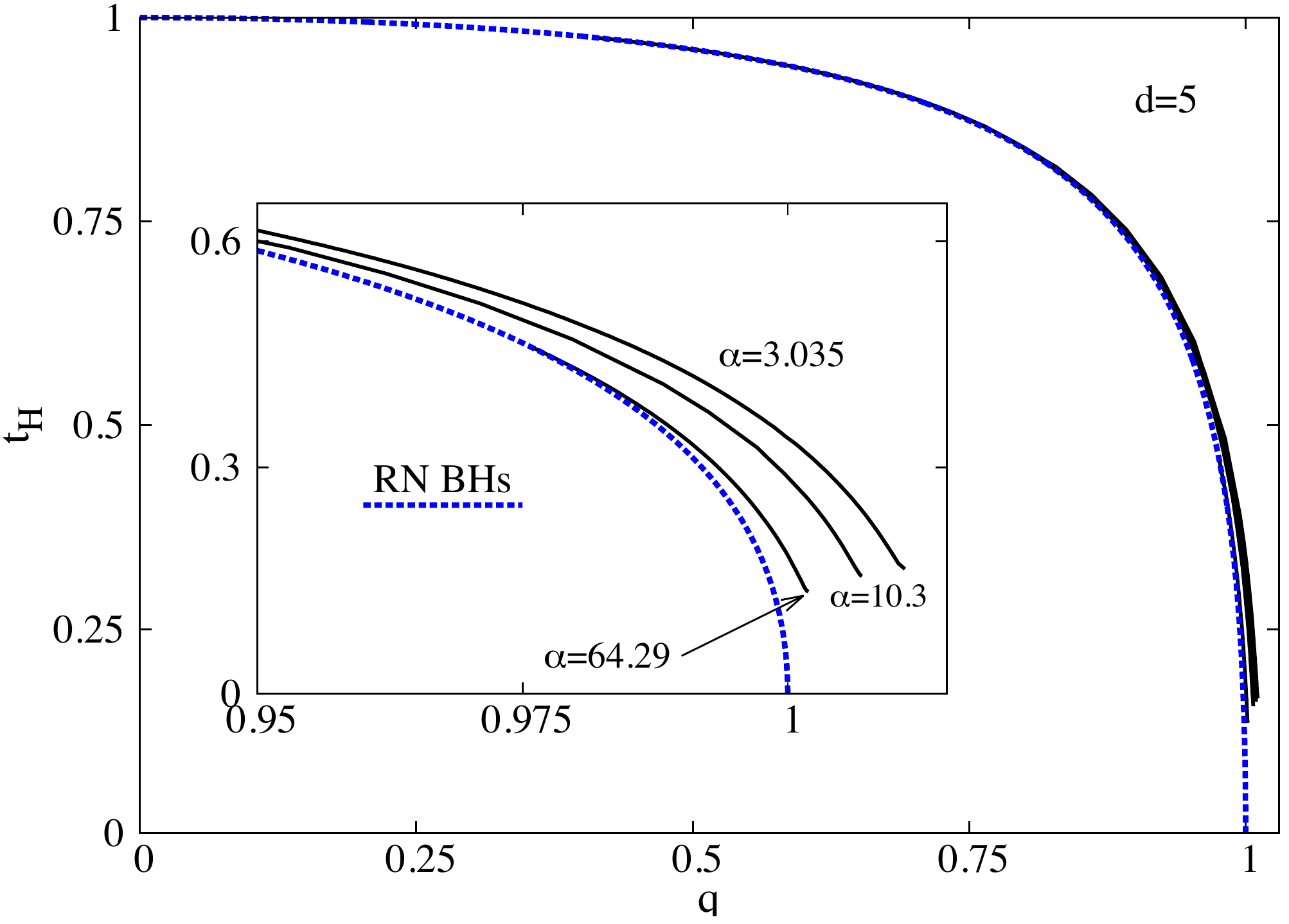} 
\caption{ 
Sequences of scalarized $d=5$ RN BHs, 
with several values of $\alpha$,  in  a charge $vs.$ horizon area (left panel)  and a charge $vs.$ Hawking temperature (right panel) diagram.
The quantities are the reduced ones, $i.e.$ given in units of mass.
}
\label{d=5}
\end{center}
\end{figure} 
Moreover, the properties of the five dimensional solutions appear to be generic.
Also, only nodeless solutions (in the scalar field) were studied so far, corresponding to the fundamental states; but solutions with nodes should also exist, corresponding to excited states.

The basic properties of the $d=5$ scalarized RN BHs can
be summarized as follows.
Given a value of the coupling constant $\alpha$, 
the spherically symmetric BHs bifurcate from the RN solution supporting the corresponding scalar cloud, as discussed in the previous subsection.
Keeping constant the parameter $\alpha$, this branch
has a finite extent,
ending in a critical configuration.
This limiting solution appears to be singular,
as found when evaluating the Kretschmann scalar at the horizon,
although its horizon area and global charges remain finite.
 This is illustrated in Fig.~\ref{d=5}, wherein the
 reduced charge $vs.$ horizon area (left panel)
and $vs.$ Hawking temperature (right panel) diagrams are exhibited, normalized $w.r.t$. the mass,  
	for several values of the coupling constant $\alpha$. As these constant $\alpha$ sequences of scalarized BHs emerge from RN BHs, the ratio $q \sim Q/M$
increases and becomes slightly larger than one, in a region close to the critical configuration where the sequence ends. In this sense, the scalarized BHs can be \textit{overcharged}, that is, they can support more charge to mass ratio than RN BHs.  
To summarize,
in an $(\alpha,q)$-diagram, the domain of existence 
of the scalarized solutions is delimited by two curves:
i) the existence line (RN BHs) and
ii) the critical line, which is the set of all critical solutions discussed above.

\subsection{Einstein frame picture and the relation to Einstein-Maxwell scalar models}
The model (\ref{actionEM}) is formulated in the so called \textit{Jordan frame},
 wherein the scalar field is non-minimally coupled to the Ricci scalar. But it possesses an equivalent formulation 
in the \textit{Einstein frame},
with a minimally coupled scalar field to the Ricci scalar but non-minimally coupled to the Maxwell invariant. 
That is, performing the conformal transformation
\begin{eqnarray} 
\label{transform1}
\bar{g}_{\mu\nu}=\Omega^{\frac{4}{d-2}} g_{\mu\nu} \ ,
\qquad
\Omega^{2}=1- \alpha \phi^2\ ,
\end{eqnarray}
together with a redefinition of the scalar field
\begin{eqnarray}
 \label{transform2}
d  \psi=\frac{\sqrt{1- \alpha [1-\frac{8\alpha (d-1)}{d-2}]\phi^2 }}{1-\alpha \phi^2 }d \phi \ ,
\end{eqnarray}
transforms  (\ref{actionEM}) into the Einstein frame action
\begin{eqnarray} 
\label{newaction}
\mathcal{S}=\frac{1}{16\pi}\int d^{d}x\sqrt{-\bar{g}}
\left[\bar{R}
   - \frac{1}{2}\bar g^{\mu\nu} \partial_\mu   \psi  \partial_\nu  \psi 
-f( \psi)
\bar g^{\mu\nu} \bar g^{\alpha\beta} F_{\mu\alpha}  F_{\nu\beta} 
\right] \  ,
\end{eqnarray}
with the coupling function
\begin{eqnarray} 
\label{fpsi}
f(\psi)=   \Omega^{-\frac{2(d-4)}{d-2}}(\psi )= \left( 1-\alpha \phi^2(\psi) \right)^{-\frac{d-4}{d-2}}   \ .
\end{eqnarray}
The new, Einstein frame, variables are the metric $\bar{g}_{\mu\nu}$ and the scalar field $  \psi$.
 The transformation 
given by eqs. (\ref{transform1}) and (\ref{transform2}) therefore maps a solution of the field 
equations 
(\ref{st1})-(\ref{st2}), 
 to a solution 
that extremizes (\ref{newaction}). The transformation is independent of any assumption of 
symmetry, and in this sense is covariant; one can easily infer that the 
transformation is one-to-one in general.   

This transformation leads to an interesting twist:
in the Einstein frame,  the spontaneous scalarization of electrovacuum BHs 
results from the nonstandard coupling of the new scalar field $\psi$ to the Maxwell term
(notice the analogy with the case in the recent work \cite{Herdeiro:2018wub}).
That is the scalarized solutions of the scalar-tensor model can be interpreted as scalarized solutions of an Einstein-Maxwell-scalar model. 

One can use this mapping to extract information about scalarization (or lack thereof) 
of the corresponding Einstein-Maxwell-scalar model. 
 The Einstein-frame scalar field, as resulting from (\ref{transform2}) reads 
\begin{eqnarray}
\label{psi-gen}
&&
\psi=-\frac{\sqrt{8(d-1)(\alpha-\alpha_c)}}{\sqrt{(d-2)\alpha}}
 {\rm arcsinh} \left( \sqrt{\frac{8(d-1)\alpha (\alpha-\alpha_c)}{d-2}} \phi\right) 
\\
\nonumber
&&
{~~~~~~~~}
+\frac{2\sqrt{2(d-1)}}{\sqrt{(d-2)}} {\rm arctanh} 
\left( 
\frac{2\sqrt{2(d-1)}\alpha \phi}{\sqrt{d-2+ 8(d-1)(\alpha-\alpha_c)\alpha \phi^2}}
\right) \ ,
\end{eqnarray}
with
\begin{eqnarray}
\alpha_c\equiv \frac{1}{8}\frac{d-2}{d-1} \ .
\end{eqnarray}
$\alpha_c$ is a special value of $\alpha$
corresponding to a $d$-dimensional {\it conformally coupled} scalar field in the Jordan frame. 
Choosing $\alpha=\alpha_c$,  
the coupling function to the Maxwell invariant in the Einstein frame can be computed in closed form, yielding
\begin{eqnarray}
f(\psi)=  {\cosh^{\frac{2(d-4)}{d-2}}  \left(\frac{1}{2} \sqrt{\frac{d-2}{2(d-1)}}   \psi \right)} \ .
\end{eqnarray}
Unfortunately, it is simple to verify that the value $\alpha=\alpha_c$ of the coupling constant does \textit{not} obey (\ref{con1s}). 
Thus, a  conformally coupled scalar will not allow the scalarization of the RN BH in the scalar-tensor model. 

However, the scalarization  becomes possible for large enough values of $\alpha$.
In fact, as  long as $\Omega^2>0$
all  solutions of the initial model (\ref{actionEM}) are mapped to  
BHs  of the Einstein frame model (\ref{newaction}).
The corresponding expression of the coupling function 
results by inverting (\ref{psi-gen}) and replacing in (\ref{fpsi}).
%
Although   $f(\psi)$
cannot be found in closed form (unless $\alpha=\alpha_c$),
its expression for   
 a small enough scalar field reads
%
\begin{eqnarray}
f(\psi) \simeq 1+\beta \psi^2+\mathcal{O}(\psi^4) \ , \qquad 
{\rm where}~~\beta\equiv \frac{ \alpha (d-4) }{d-2}~.
\end{eqnarray}
One remarks that $d=4$ scalarized RN BHs with the above form of the coupling function
have been studied in 
\cite{Fernandes:2019rez,Boskovic:2018lkj}, and 
they capture the basic properties of the generic case.

\section{Scalarized vacuum BHs in $d>4$ extended scalar-tensor  models}
\label{sec4}

\subsection{The framework}
For our second sub-class of models we consider an \textit{extended} scalar-tensor model. Thus, we take (\ref{actiongen}) with (\ref{guise2}) and $\mathcal{L}_{\rm mat}=0$.  The explicit expressions of the first terms in the hierarchy of $\mathcal{L}_{(p)}$ are
\begin{eqnarray}
&& \mathcal{L}_{(0)} = 1 \ , \qquad 
~~\mathcal{L}_{(1)} = R \ , \qquad 
~~
\mathcal{L}_{(2)} = R^2-4R_{\mu\nu}R^{\mu\nu}+R_{\mu\nu\rho\sigma}R^{\mu\nu\rho\sigma} \ ,
\\
&&
\mathcal{  L}_{(3)} = R^3   -12RR_{\mu \nu } R^{\mu \nu } + 16R_{\mu \nu }R^{\mu }_{\phantom{\mu } \rho }R^{\nu \rho }+ 24 R_{\mu \nu }R_{\rho \sigma }R^{\mu \rho \nu \sigma }+ 3RR_{\mu \nu \rho \sigma } R^{\mu \nu \rho \sigma } 
\nonumber 
\\
&&{~~~~~~~~~}
-24R_{\mu \nu }R^\mu _{\phantom{\mu } \rho \sigma \kappa }
 R^{\nu \rho \sigma \kappa  }+ 4 R_{\mu \nu \rho \sigma }R^{\mu \nu \eta \zeta } 
R^{\rho \sigma }_{\phantom{\rho \sigma } \eta \zeta }
-8R_{\mu \rho \nu \sigma } R^{\mu  \phantom{\eta } \nu }_{\phantom{\mu } \eta 
\phantom{\nu } \zeta } R^{\rho  \eta  \sigma  \zeta } \ .
\end{eqnarray}

In constructing higher dimensional generalizations of the scalarized BHs in~\cite{Silva:2017uqg,Doneva:2017bvd,Antoniou:2017acq},
we use the observation that,
in even dimensions, the contribution to the action of the $d/2$-th order $\mathcal{L}_{(p)}$ becomes
a topological invariant, and alone   does not contribute 
to the field equations.
This ceases to be the case when a nontrivial coupling function, $f(\phi)$ is present:
the term
$\mathcal{  L}_{(d/2)} $ becomes dynamical.
As an example, for  $d=4$ one takes $p=2$
($i.e.$ the GB term) 
and
the geometrical scalarization model in \cite{Silva:2017uqg,Doneva:2017bvd,Antoniou:2017acq} 
is recovered. 

In what follows, we investigate solutions of the model (\ref{actiongen}) with (\ref{guise2}), $\mathcal{L}_{\rm mat}=0$ and
\begin{eqnarray}
d=2p\ , \qquad {\rm where} \qquad p\geqslant 2 \ ,
\end{eqnarray}
and show that the properties of the four dimensional solutions are generic. As in the previous section, for ease of notation we drop the subscript label in the coupling constant: $\alpha_{\rm L}\rightarrow \alpha$.
Thus, the considered action reads
\be
\label{actionL}
\mathcal{S}=- \frac{1}{16 \pi}\int d^{2p} x \sqrt{-g} 
\left\{
R
-\frac{1}{2}\partial_\mu  \phi \partial^\mu \phi 
+\alpha  f (\phi) \mathcal{L}_{(p)} 
\right\} \ .
\ee

\subsection{The equations of motion and general results}
In obtaing the equations of motion it is useful to observe that, for general $p$, the variation of the Euler density term is
\be
\frac{\delta(f (\phi)\mathcal{  L}_{(p)})}{\delta g^{\mu\nu}} = -2p P^{(p)}_{\mu\rho\nu\beta}\nabla^\rho\nabla^\beta f(\phi) \ ,
\ee
where the $P^{(p)}_{\mu\nu\alpha\beta}$ tensor is naturally defined in $2p$ dimensions using the Levi-Civita tensor in that dimension:
\be
P^{(p)\mu\nu\alpha\beta} =-\frac{1}{2^p}\epsilon^{\mu\nu\mu_1\nu_1...\mu_{p-1}\nu_{p-1}}\epsilon^{\alpha\beta\alpha_1\beta_1...\alpha_{p-1}\beta_{p-1}}R_{\mu_1\nu_1\alpha_1\beta_1}...R_{\mu_{p-1}\nu_{p-1}\alpha_{p-1}\beta_{p-1}} \ .
\ee
We remark that this tensor shares some of the symmetries and properties of the Riemann tensor:
\be
P^{(p)}_{\mu\nu\alpha\beta} = - P^{(p)}_{\nu\mu\alpha\beta} = - P^{(p)}_{\mu\nu\beta\alpha} \ ,
\qquad 
P^{(p)}_{\mu\nu\alpha\beta} = P^{(p)}_{\alpha\beta\mu\nu} \ , 
\qquad 
\nabla^\mu P^{(p)}_{\mu\nu\alpha\beta} = 0 \ .
\ee

Taking the same ansatz as before for the metric and scalar field (\ref{metric-scalar}), a straightforward (but cumbersome) computation leads to the following
 equations for the metric functions and the scalar field\footnote{There is yet another second order equation, which,
however, can be expressed as a linear combination of Eqs. (\ref{eqH}), (\ref{eqs}) and their first derivatives,
together with (\ref{eqf}). }
\begin{eqnarray} 
\label{eqH}
&&
(d-2)N'
-(d-2)(d-3) \frac{(1-N)}{r}
+\frac{r}{2} N\phi'^2
+\frac{2\alpha}{r^{d-3}}(1-N)^{\frac{d-4}{2}}
\\
\nonumber
&&
{~~~~~~~~~~~~~~~~~~~~~~}
\times
\left[
(1-N)N f''(\phi)-\frac{1}{2}\left\{(d-1)N-1\right\}N'f'(\phi)
\right]=0 \ ,
\\ 
&&
\label{eqs}
(d-2)\sigma'
-  \frac{r}{2} \sigma\phi'^2
+\frac{ \alpha}{r^{d-3}}(1-N)^{\frac{d-4}{2}}
\left[
(1-N)\sigma f''(\phi)+\left\{(d-1)N-1\right\}\sigma'f'(\phi)
\right]=0\ , \ \ \ \ \ \ \ \ 
\\
&&
\label{eqf}
(N\sigma r^{d-2}\phi')'-\alpha \frac{df(\phi)}{d\phi} \left\{
(1-N  )^{\frac{1}{2}(d-2)}
\left (
\sigma N'+2N \sigma'
\right)
\right\}'=0 \ .
\end{eqnarray}
These equations can also be derived from the
effective Lagrangian\footnote{Here, as well as in the equations
(\ref{eqH})-(\ref{eqf}),
to simplify the relation,
we have absorbed
in
$\alpha$
 a factor of
$\frac{1}{2}(d-2)!$
.
}
\begin{eqnarray}
{\cal L}_{\rm eff}={\cal L}_{\rm E}+{\cal L}_{s}+ \alpha {\cal L}_{(p)} \ ,
\end{eqnarray}
with  ${\cal L}_{\rm E}$ and $ {\cal L}_{s}$ given by (\ref{lels}) and 
\begin{eqnarray} 
 {\cal L}_{(p)}=\frac{dT_{(p)}}{dr} \ ,
\qquad 
 T_{(p)}\equiv -(1-N  )^{\frac{1}{2}(d-2)}
\left (
\sigma N'+2N \sigma'
\right) \ .
\end{eqnarray}

As in the last section, we are interested in BH solutions,
with a horizon at $r=r_h>0$.
Restricting to non-extremal configurations,
the near horizon expansion of the solutions reads
\begin{eqnarray}
N(r)=N_1(r-r_h)+\dots\ , \qquad \sigma(r)=\sigma_h+ \sigma_1 (r-r_h)+\dots \ , \qquad 
\phi(r)=\phi_h+\phi_1 (r-r_h) +\dots \ .
\end{eqnarray}
All coefficient are determined by the essential parameters
$r_h$,
$\phi(r_h)$
and $\sigma(r_h)$;
for example, one finds
\begin{eqnarray}
 N_1 =\frac{(d-2)(d-3)}{r_h[d-2_\alpha\phi_1 f'(\phi_h)]} \ .
\end{eqnarray}
The coefficient 
$\phi'(r_h)$
satisfies a  second order algebraic equation of the form
\begin{eqnarray}
\label{eq2}
\phi_1^2+p \phi_1+q=0 \ ,
\end{eqnarray}
where $(p,q)$ are non-trivial functions of $r_h,\phi_h$.
Consequently,  a real solution for $\phi_1$ of~(\ref{eq2}) exists only if $\Delta=p^2-4q \geqslant 0$,
a condition which translates into the following inequality 
\begin{eqnarray} 
\label{cond}
 1-\left( \frac{\alpha}{r_h^{d-2}} \right)^2\frac{4(d-1)(d-3)}{d-2} \left(\frac{df(\phi)}{d\phi}\bigg|_{\phi_h} \right)^2
\left\{
 1-\left( \frac{\alpha}{r_h^{d-2}} \right)^2\frac{d(d-4)}{4(d-2)}  \left(\frac{df(\phi)}{d\phi}\bigg|_{\phi_h} \right)^2
\right\}
>0 \ ,
\end{eqnarray}
which implies the existence of a minimal horizon size, 
denoted as $r_{h}^{\rm (min)}$,
determined by $\alpha$
and the value of the scalar field at the horizon.

The far field expansion of the solutions reads
\begin{eqnarray} 
\label{inf21}
N(r)=1-\frac{m}{r^{d-3}} +\dots\ , \qquad \sigma(r)=1-\frac{(d-3)}{4(d-2)}\frac{Q_s^2}{r^{2(d-3)}}+\dots \ , \qquad 
\phi(r)=\frac{Q_s}{r^{d-3}}+\dots \ ,
\end{eqnarray}
in terms of two constants: the scalar 'charge', $Q_s$, and $m$, which fixes the ADM mass $M$ as in (\ref{QMgen}).

The horizon data, corresponding to the Hawking temperature and horizon area are still given by (\ref{THAH}) (with vanishing electric charge).
 In terms of all these quantities, the solutions satisfy the 
Smarr-like relation:
\begin{eqnarray} 
M=\frac{(d-2)}{(d-3)}T_H S+M_{(\phi)} \ ,
\end{eqnarray}
where $S$ is the BH entropy as computed from Wald's formula \cite{Wald:1993nt}
\begin{eqnarray} 
S=S_{\rm EH}+S_{(p)}\ , \qquad S_{\rm EH}=\frac{1}{4}A_H\ , \qquad S_{(p)}=\frac{1}{4}\alpha V_{{d-2}}f(\phi(r_h)) \ ,
\end{eqnarray}
and $M_{(\phi)}$ is the mass stored in the scalar field
\begin{eqnarray} 
M_{(\phi)}=-\frac{d-2}{d-3}\frac{1}{16\pi}\int_\Sigma d^{d-1}x \sqrt{-g}\frac{f(\phi)}{\dot f(\phi)}\Box \phi \ ,
\end{eqnarray}
where the integral is taken over a spacelike surface $\Sigma$ and $\dot{f}\equiv df/d\phi$.

We define the \textit{reduced} horizon area and Hawking temperature as in (\ref{scale1})
by normalizing the corresponding quantities $w.r.t.$
the total mass of solutions. Analogously, the reduced entropy is defined as:
 \begin{eqnarray} 
 s\equiv \frac{4S}{M^{\frac{d-2}{d-3}}}c_a\ ,
\end{eqnarray}
where $c_a$ is given by (\ref{scale2})

The scalar-free solution in this model is the Schwarzschild-Tangherlini BH
\cite{Tangherlini:1963bw}, which has 
a vanishing scalar field, 
$m=r_h^{d-3}$, $\sigma=1$,
while its reduced quantities are simply $a_H=s=t_H=1$.

\begin{figure}[h!]
\begin{center}
\includegraphics[width=0.45\textwidth]{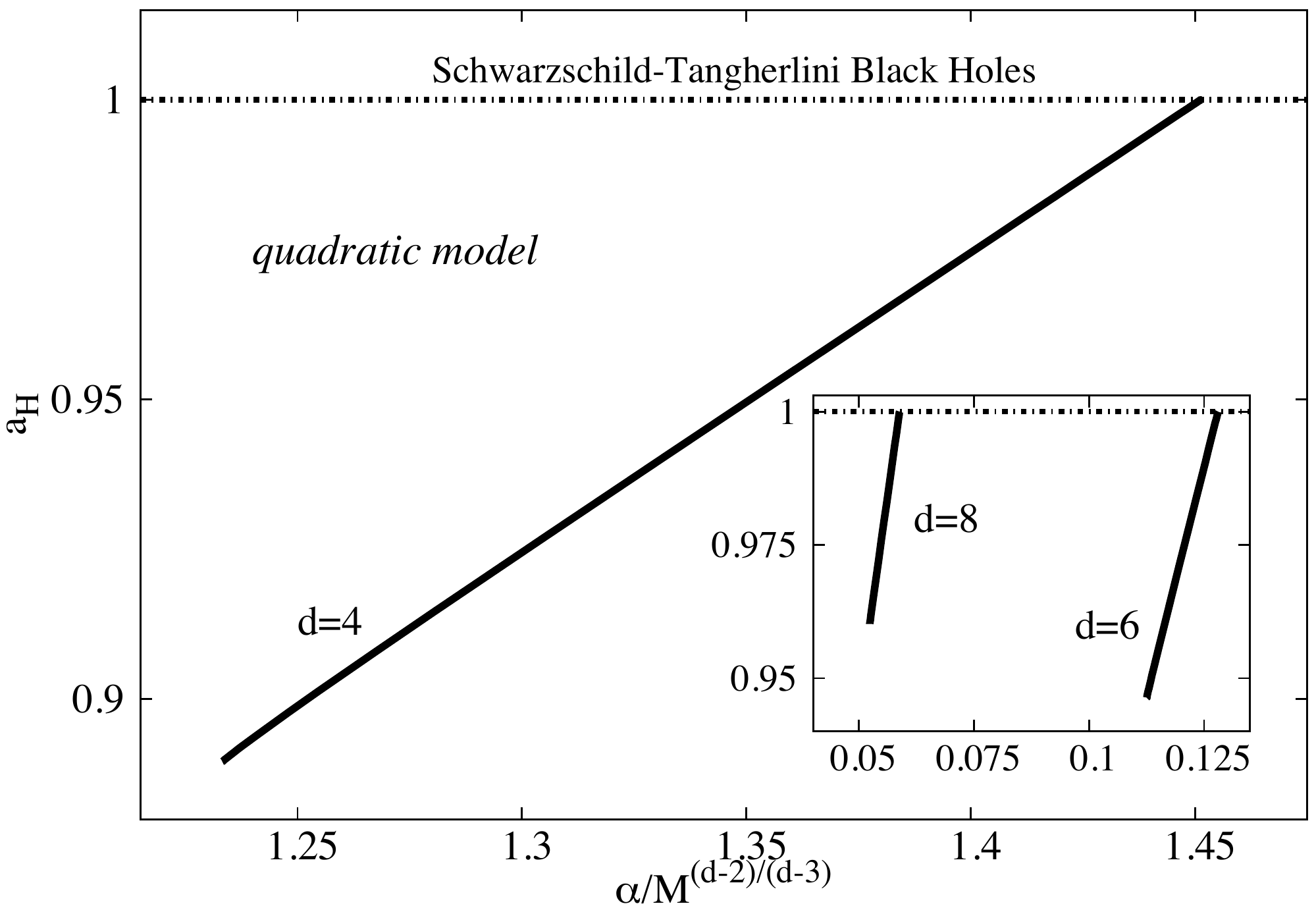}\ \ \
\includegraphics[width=0.45\textwidth]{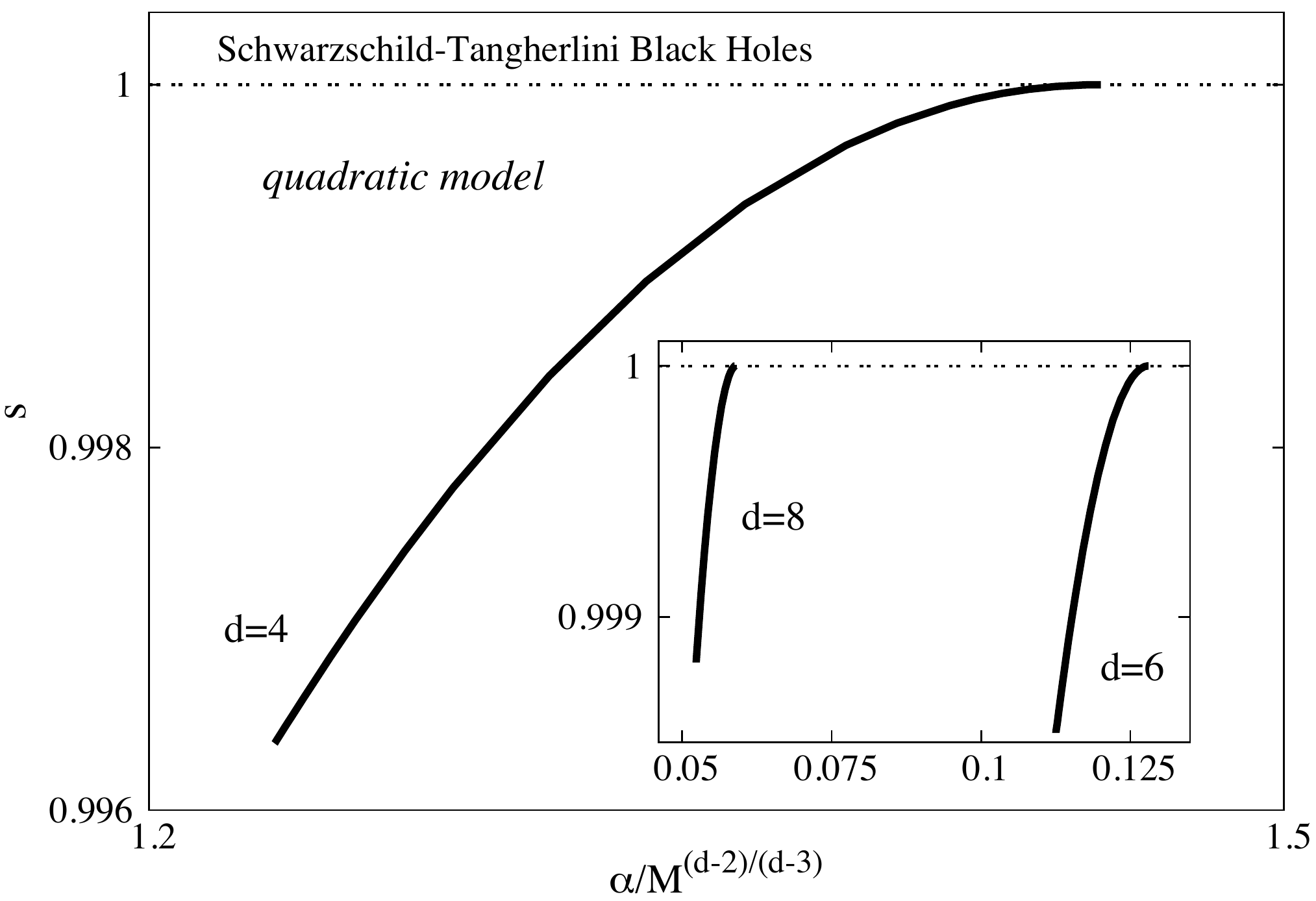} 
\includegraphics[width=0.45\textwidth]{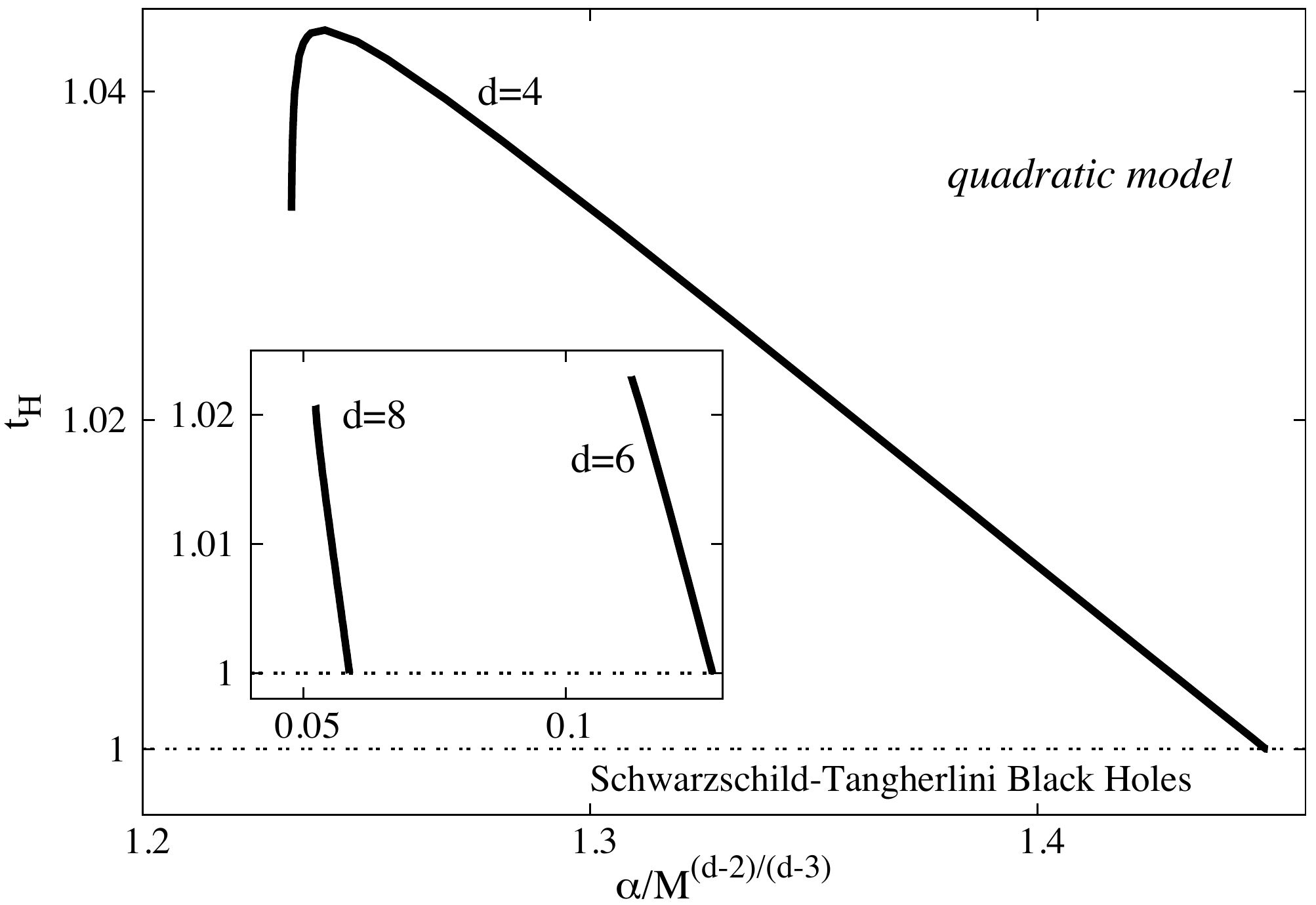}\ 
\includegraphics[width=0.45\textwidth]{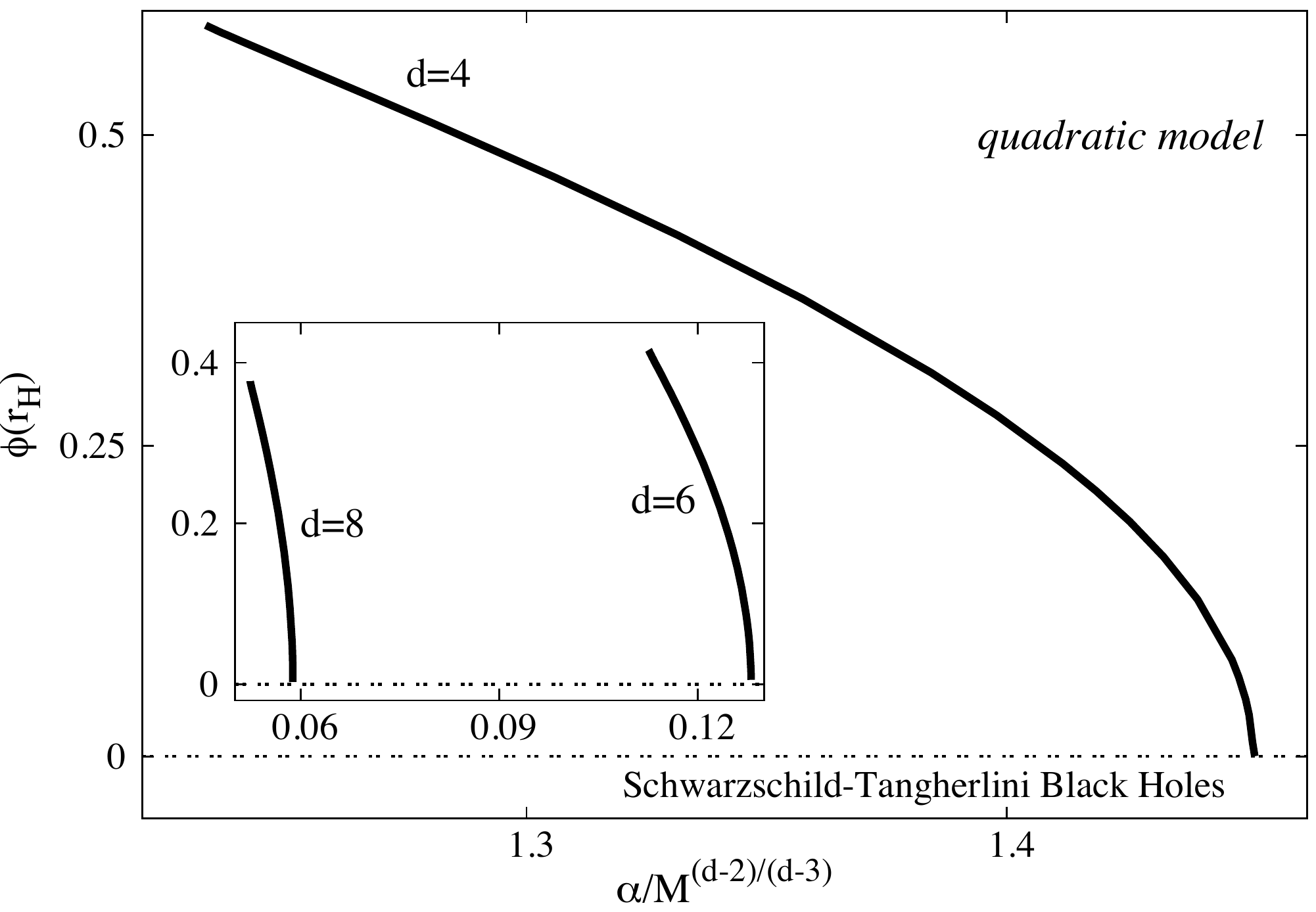}
\caption{Reduced area (top left panel), reduced entropy (top right panel), reduced temperature (bottom left panel) and 
 the scalar field at the horizon (bottom right panel) $vs.$ the coupling (normalized by the mass)
for scalarized BHs in the extended scalar-tensor model in $d=4,6,8$
dimensions.
}
\label{geometric}
\end{center}
\end{figure}  

\subsection{The scalarized BHs in $d=4,6,8$ with a quadratic coupling}
As in the previous section we shall illustrate the BHs in the $d$-dimensional extended scalar-tensor models by considering the simplest function which satisfies the condition (\ref{condx}):
\begin{eqnarray} 
f_{\rm L}(\phi)= \phi^2~,
\label{sq}
\end{eqnarray}
which was initially considered for  $d=4$ solutions in~\cite{Silva:2017uqg}. The numerical construction of the solutions in $d=4,6,8$ follows a strategy similar to one used in the last section. Some properties of the solutions are shown in Fig.~\ref{geometric} and can be summarized as follows. 
Firstly,  the qualitative features of the $d=4$
solutions still hold in higher $d$, namely: (i) the branching off from the Schwarzschild-Tangherlini BH wherein the latter supports a scalar cloud; (ii) the limited range wherein solutions exist; (iii) and the trends of the different quantities when $\alpha$ is varied. Quantitatively, however, one can see a smaller domain of existence in terms of $\alpha$ in higher dimensions, likely due to the faster fall-off of the gravitational interaction.
Secondly, the model possesses a (presumably infinite) tower of scalarized spherically symmetric solutions which are
 indexed by the number of nodes  $n$  of the scalar field.
As in the previous study of the scalar-tensor model, here we are focusing on the fundamental $n=0$ solutions.
Thirdly, all solutions can be obtained continuously in the parameter space: they
form a line, starting from the smooth GR limit ($\phi\to 0$), and ending at some limiting solution. 
Once the limiting configuration is reached, the solutions cease to exist in
the parameter space.
The existence of these 'critical'
configurations  can be understood from the condition
(\ref{cond}), with the determinant $\Delta$ vanishing at that point.
%

\subsection{A linear coupling detour: the shift-symmetric model in $d$-dimensions}
 If instead of the choice~(\ref{sq}) one chooses the coupling function
\begin{eqnarray} 
\label{linear}
f_{\rm L}(\phi)=\phi \ ,
\end{eqnarray}
the scalarization condition (\ref{condx})  is not obeyed. 
This case
corresponds to a linear coupling or a 'shift symmetric'
model, which is interesting for different reasons and has been extensively studied for $d=4$ - see $e.g.$  \cite{Benkel:2016rlz,Sotiriou:2014pfa,Sotiriou:2013qea,Delgado:2020rev}. Although
 scalarization is absent,
the model possesses a variety of interesting properties. Here, we shall use it to contrast with the picture found for the quadratic coupling 
in the previous subsection.

Since the condition (\ref{condx})
is not satisfied in the linear model (\ref{linear}) for $\alpha \neq 0$,
the Schwarzschild-Tangerlini BH is not a solution.
Also,
 the equations of motion are invariant under the
transformation
\begin{eqnarray} 
 \phi \to \phi+\phi_0 \ ,
\end{eqnarray}
with $\phi_0$
an arbitrary constant, which results from the fact that  
the $\mathcal{L}_{(p)}$ term alone is a total divergence.
This implies the existence of a current,
whose conservation leads to the following
interesting relation between the 
'scalar charge' and the Hawking temperature
\begin{eqnarray} 
Q_s=\frac{4\pi \alpha}{(d-3)}T_H \ ,
\end{eqnarray}
which is a unique property of this class  of models
(see also the discussion in \cite{Prabhu:2018aun}
 for $d=4$).

 In the 
probe limit, that is considering the scalar field equation of the model on the Schwarzschild-Tangherlini background, we find the following general exact solution\footnote{Note that a constant of integration has been 
fixed in the expression by imposing  $\phi(r)\to 0$ as $r\to \infty$}:
\begin{eqnarray} 
\label{phi-probe}
\phi(r)= \frac{\alpha}{r_h^{d-2}}
\left\{
B\left[\left(\frac{r_h}{r}\right)^{d-3}; \frac{d^2-d-4}{2(d-3)},0 \right]
+\log\left(1-\left[\frac{r_h}{r}\right]^{d-3} \right)
\right\} \ ,
\end{eqnarray}
where $B[x;a,b]$
is the incomplete $\beta$-function.
Simple expressions exist for $d=4,6$ only
(with $x\equiv {r_h}/{r}$):
\begin{eqnarray} 
&&
d=4: \qquad 
\phi(r)= \frac{\alpha}{r_h^{2}}
\left(
x+\frac{x^2}{2}+\frac{x^3}{3}
\right) \ ,
\\
&&
d=6: \qquad 
\phi(r)= \frac{3\alpha}{r_h^{4}}
\left[
x+\frac{x^4}{4}+\frac{x^7}{7}+\frac{x^{10}}{10}\
-\frac{1}{2}\log(1+x+x^2)
-\frac{1}{\sqrt{3}}\arctan\left(\frac{\sqrt{3}x}{2+x}\right)
\right] \ .
\end{eqnarray}
In principle, this solution can be used
to construct a closed form
 perturbative solution as a power series
in the parameter $\alpha/r_h^{d-3}$,
see $e.g.$
the $d=4$ results in \cite{Delgado:2020rev}.
As discussed therein, this  analytical solution 
 provides  a  good approximation to the
numerical results.

A feature which, however, cannot be captured within a perturbative approach 
is the existence of 
a minimal horizon size.
The condition (\ref{cond})
on the near horizon data takes a simple form for the choice 
(\ref{linear}) of the coupling function, with
\begin{eqnarray} 
\label{condn}
 \frac{\alpha}{r_h^{d-2}}<
\left[
\frac{2(d-1)(d-3)}{(d-2)}+\sqrt{3(d-1)(d-3)}
\right]^{-1/2}.
\end{eqnarray}
This  requirement translates into a  coordinate independent condition imposing a minimal size for
the horizon
size in terms of  the coupling constant $\alpha$ only,
\begin{eqnarray} 
A_H > c_0 \alpha \ , \qquad {\rm where} \qquad c_0\equiv  V_{(d-2)}\sqrt{\frac{2(d-1)(d-3)}{(d-2)}+\sqrt{3(d-1)(d-3)}} \ .
\end{eqnarray}

\begin{figure}[h!]
\begin{center}
\includegraphics[width=0.45\textwidth]{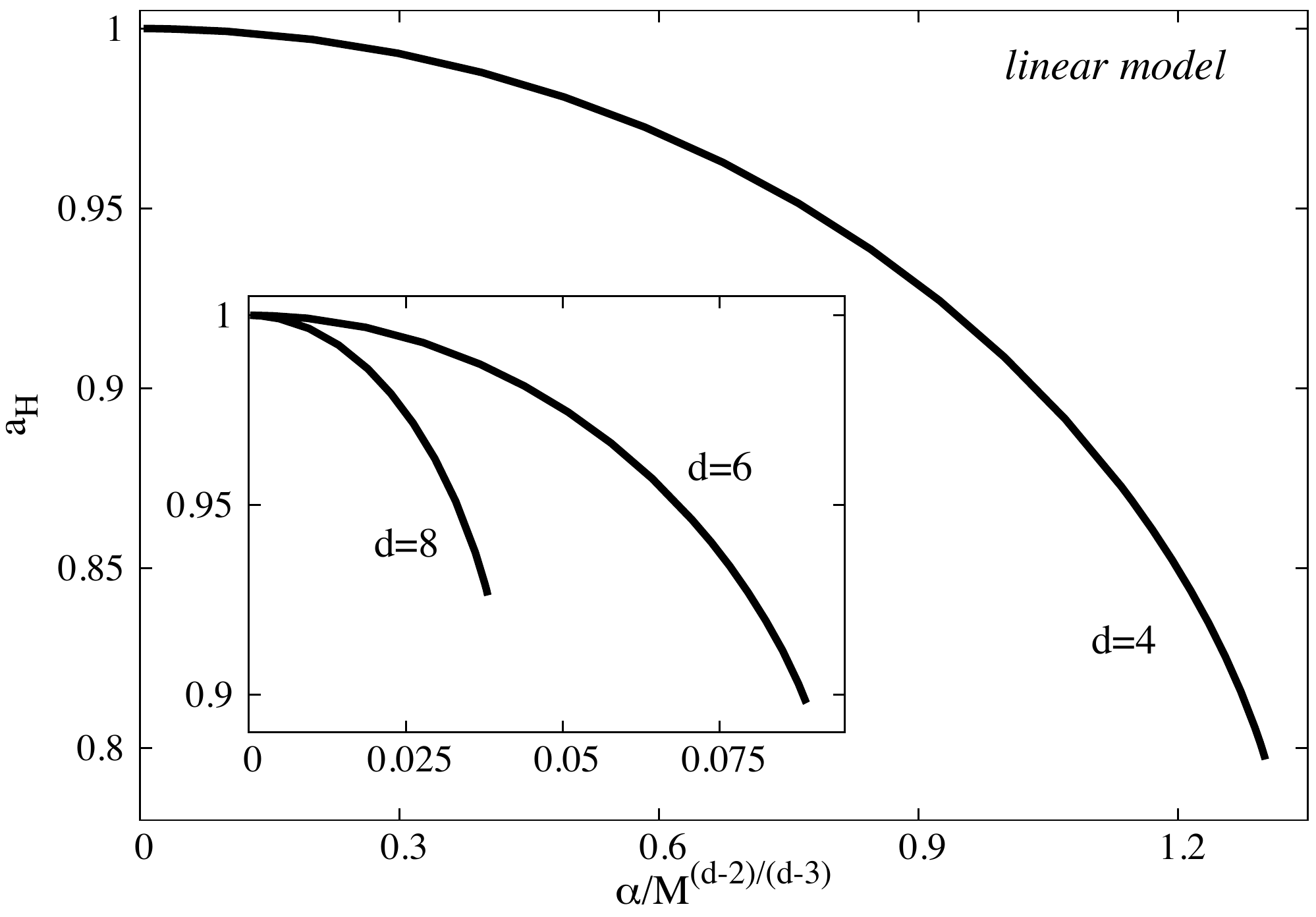}\ \ \
\includegraphics[width=0.45\textwidth]{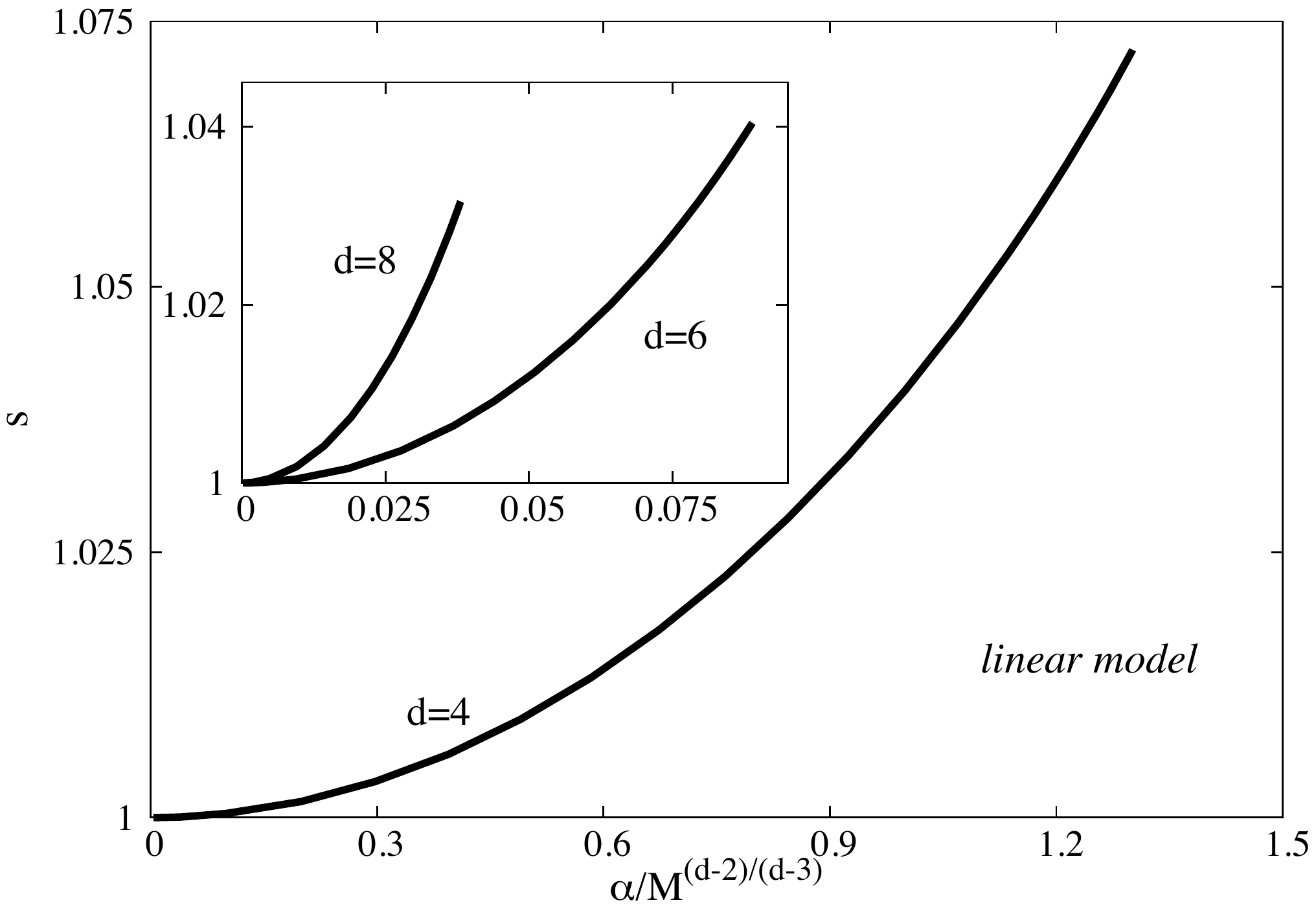} 
\includegraphics[width=0.45\textwidth]{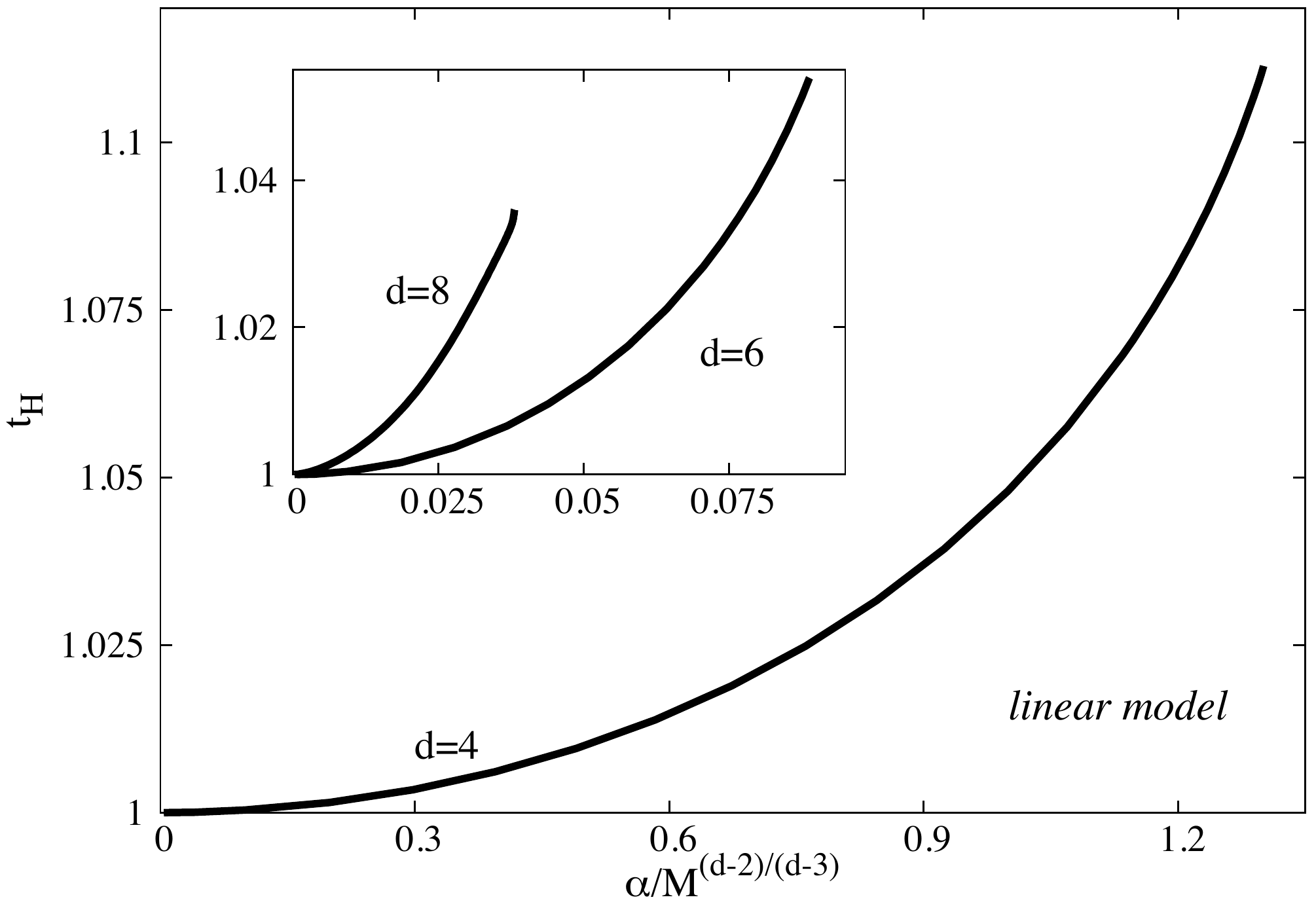}\ 
\includegraphics[width=0.45\textwidth]{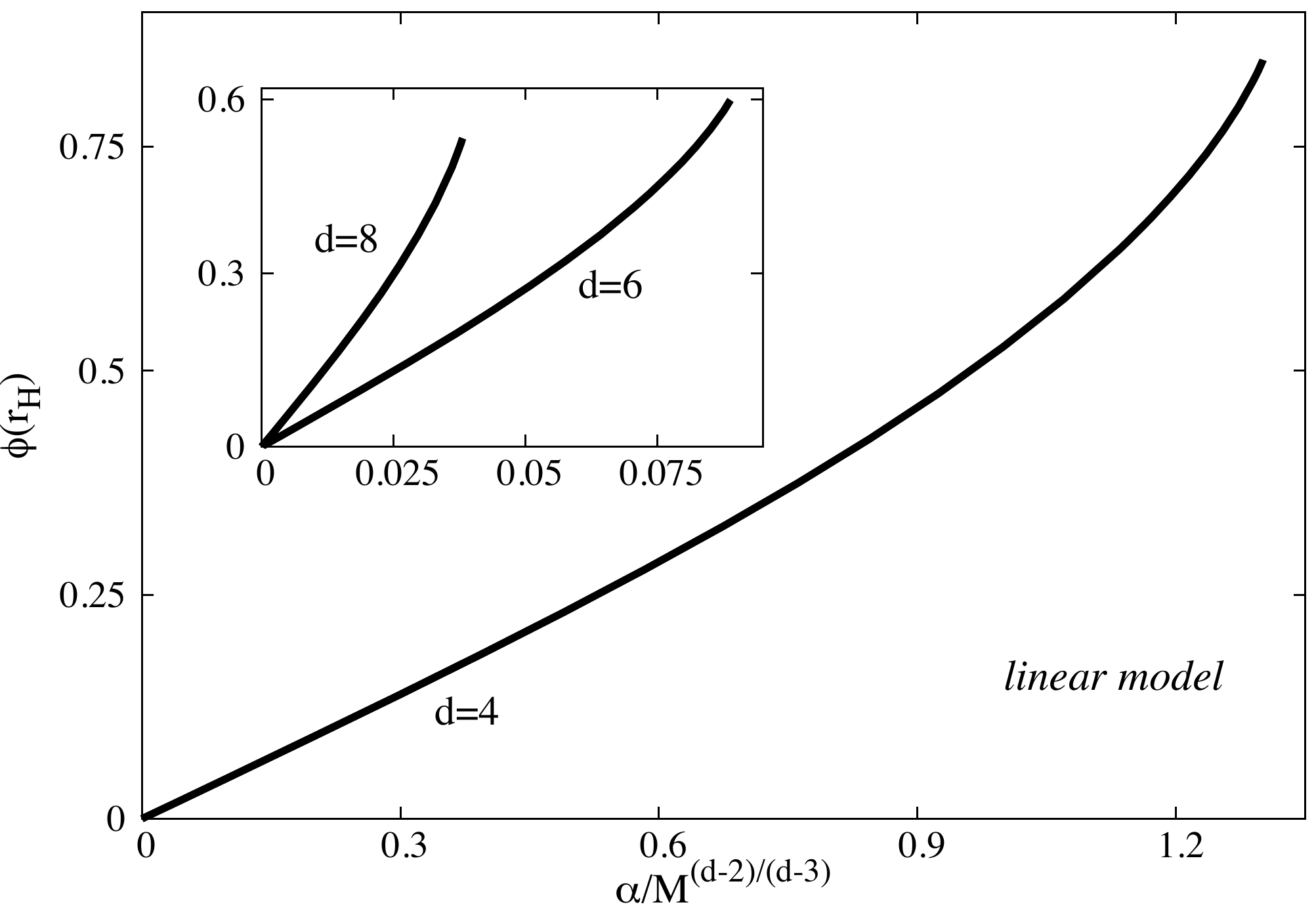}
\caption{Reduced area (top left panel), reduced entropy (top right panel), reduced temperature (bottom left panel) and 
 the scalar field at the horizon (bottom right panel)
for BHs
in the shift symmetric model 
 in $d=4,6,8$ spacetime
dimensions.}
\label{linear1}
\end{center}
\end{figure}

Some results of the numerical integration for non-perturbative solutions 
are shown in Fig.~\ref{linear1}.
Again,
the solutions stop existing at the point where the condition (\ref{condn})
fails to be satisfied.

\section{Summary and overview}
\label{sec5}

The main purpose of this paper was to discuss higher dimensional generalizations of $d=4$
spontaneous scalarizations models, in its various guises, via the existence of the corresponding scalarized BH solutions. 
As the broader take home message, the study herein shows the phenomenon of 'spontaneous scalarization'
is not peculiar to $d=4$, but qualitative and quantitative differences occur in higher $d$.\footnote{In various aspects $d=4$ 
BH physics has unique properties; 
recent research has revealed that as $d$ increases, the BH's phase structure becomes
increasingly intricate and diverse \cite{Emparan:2008eg}. This further motivates the analysis herein.}

Concerning the case of the scalar-tensor model studied in Section~\ref{sec3}, 
we have established that, since the conformal invariance of the Maxwell action is lost in $d>4$,
 the higher dimensional electrovacuum  BHs
possess scalarized generalizations in these models. This is a qualitative difference with respect to the $d=4$ case. Moreover, by a conformal mapping, these solutions can be interpreted as Einstein-Maxwell-scalar solutions, bridging between these two guises of scalarization.

Concerning the case of the extended scalar-tensor model studied in Section~\ref{sec4}, 
our construction generalized the  'geometric scalarization' in~\cite{Silva:2017uqg,Doneva:2017bvd,Antoniou:2017acq}
to any $even$ dimension. In $d=4$, 
Einstein's gravity can be deduced by assuming general coordinate covariance and the
absence of higher derivative terms larger than the second order in the Lagrangian.
 In $d>4$ , the same assumptions lead to Lovelock gravity	
\cite{Lovelock:1971yv}.	
All Euler densities, $\mathcal{L}_{(p)}$, starting with the Ricci scalar and the GB curvature squared combination, 
can be written as the divergences of genuine vector densities in the critical dimensions $d=2p$,
with $p=1,2,\dots$
(while they vanish for $d<2p$).
However, 
such a density can be made dynamical by coupling it to a scalar field,
which results in the term $\alpha_{\rm L} f_{\rm L}(\phi) \mathcal{L}_{(p)}$ in the action (\ref{actiongen}).
Thus, there is a hierarchy of models, with the $d=4$ ($p=2$)
case in 
\cite{Silva:2017uqg,Doneva:2017bvd,Antoniou:2017acq}
being a special case.
Here, we have found that the properties of the solutions of the latter 
are generic, being shared by the higher dimensional $d=2p$ counterparts, but with quantitative differences.

As to provide a comparative benchmark, we have also generalized the $d=4$ `shift symmetric'
Horndeski model in 
\cite{Benkel:2016rlz,Sotiriou:2014pfa,Sotiriou:2013qea}
to any $d=2p\geqslant 4$ even dimension.
Again, the properties of the $d=4$ solutions are generic.
Although these configurations do not qualify for scalarized BHs
(in particular  the condition (\ref{condx}) is not satisfied),
they possess a variety of interesting properties, most notably that the Hawking temperature is fixed by the scalar charge.

All configurations in this work are spherically symmetric and asymptotically flat, being regular on and outside the horizon (which possesses a spherical
topology). Rotating generalizations should exist, following the $d=4$ studies in, 
$e.g.$~\cite{Delgado:2020rev,Cunha:2019dwb,Collodel:2019kkx}.

Let us close this work with some remarks concerning the status of the extended scalar-tensor model for the  
(lower) dimension $d=2$. 
Einstein gravity alone is trivial 
in two dimensions;
however, as in the generic $d=2p$ 
case,
 ${\cal L}_{(1)}=R$ can contribute to
the equations of motion
by coupling it with a scalar field.
This suggest to consider the following $d=2$ version of the generic model (\ref{actiongen}):
\be
\label{action2}
\mathcal{S}=- \frac{1}{16 \pi}\int d^2 x \sqrt{-g} 
\left\{
\alpha f(\phi) R
-\frac{1}{2}\partial_\mu  \phi \partial^\mu \phi  
+U(\phi)
\right\} \ ,
\ee
with $U(\phi)$ a scalar potential.
Interestingly, this corresponds to the generic form of the
  Jackiw-Teitelboim (JT) gravity \cite{Jackiw:1984je, Teitelboim:1983ux}.
 This model has received considerable interest recently in connection with BH dynamics
 (see, $e.g.$ 
 \cite{Almheiri:2014cka,Castro:2019crn, Nayak:2018qej, Moitra:2019bub}). 
Near extremal BHs/branes have a near horizon `throat region' corresponding to an $AdS_2$ spacetime 
\cite{Astefanesei:2006dd} and so, upon compactification, the action (\ref{action2})  
appears naturally, with the scalar field representing the modulus associated to the transverse directions 
(the volume of the transverse sphere). 
Moreover, it was shown \cite{Moitra:2019bub} that the JT model 
 is  a  good  approximation for the low-temperature dynamics and thermodynamics  
of a large class of spinning/charged BHs, including the near extremal Kerr BH.  
  It would be interesting to  study solutions of the model 
	(\ref{action2}) for various choices of the coupling function,
	in particular for a 
	$f(\phi)$ allowing for scalarization.

\section*{Acknowlegements}

The research of DA is supported by the Fondecyt Grants 1200986, 1170279, 1171466, and 2019/13231-7  Programa de Cooperacion Internacional, ANID. J.O. is supported by the FCT grant PD/BD/128184/2016.  This work is supported by the Center for Research and Development in Mathematics and Applications (CIDMA) and by the Centre of Mathematics (CMAT) through the Portuguese Foundation for Science and Technology (FCT - Fundacao para a Ci\^encia e a Tecnologia), references UIDB/04106/2020, UIDP/04106/2020, UIDB/00013/2020 and UIDP/00013/2020, and by national funds (OE), through FCT, I.P., in the scope of the framework contract foreseen in the numbers 4, 5 and 6 of the article 23, of the Decree-Law 57/2016, of August 29, changed by Law 57/2017, of July 19.  We acknowledge support  from  the  projects  PTDC/FIS-OUT/28407/2017  and  CERN/FIS-PAR/0027/2019.   This work has further been supported by the European Union’s Horizon 2020 research and innovation (RISE) programme H2020-MSCA-RISE-2017 Grant No. FunFiCO-777740.  The authors would like to acknowledge networking support by the COST Action CA16104.


	
\end{document}